\documentstyle[12pt]{article}

\input{tcilatex}
\begin{document}

\begin{flushright}

\end{flushright}
\vskip .1in

\begin{center}
{\Large Minimal Flavour Mixing of Quarks and Leptons}

{\sf \vspace{50pt}}

{\bf J.L. Chkareuli}

\vspace{6pt}

{\em Institute of Physics, Georgian Academy of Sciences, 380077 Tbilisi,
Georgia\\[0pt]
}

\vspace{18pt}

{\large {\bf Abstract}}

\medskip
\end{center}

{Present situation in the flavour mixing of quarks and leptons is briefly
reviewed and a new approach called the Minimal Flavour Mixing (MFM) is
considered in detail. According to MFM the whole of the flavour mixing is
basically determined by the physical \ \ mass \ \ generation of \ the first
\ family of \ fermions. So, in the chiral symmetry limit when the masses of
the lightest quarks, $u$ and $d$, vanish, all the weak mixing angle vanish.
This minimal pattern is shown to fit extremely well the already established
CKM matrix elements and to give fairly distinctive predictions for the as
yet poorly known ones. Remarkably, together with generically small quark
mixing it also leads to the large neutrino mixing thus giving adequate
solution to the solar and atmospheric neutrino oscillation problem. The
possible origin of this approach\ in \ the\ MSSM extended by \ the \ \
high-scale $\ SU(3)_{F}$ chiral family symmetry is discussed.\ } \vspace{%
100pt}

\begin{center}
{\em Plenary talk given at the SUSY'01: 9th International Conference }

{\em on Supersymmetry and Unification of Fundamental Interactions }

{\em (11-17 June 2001, JINR, Dubna, Russia)}
\end{center}

\thispagestyle{empty} \newpage

\section{Preamble}

The flavour mixing of quarks and leptons is certainly one of the major
problems that presently confront particle physics. Many attempts have been
made to interpret the pattern of this mixing in terms of various family
symmetries---discrete or continuous, global or local (for a recent review
see \cite{review}). However, despite some progress in understanding the
flavour mixing problem, one has the uneasy feeling that, in many cases, the
problem seems just to be transferred from one place to another. The peculiar
quark-lepton mass hierarchy is replaced by a peculiar set of flavour charges
assigned to fermions or a peculiar hierarchy of the extra Higgs field VEVs.
As a result, apart from an original Cabibbo mixing there are not in fact
distinctive and testable generic predictions concerning the flavour mixing
angles, especially those for the heaviest quarks $t$\ and $b$. Meanwhile,
the present observational status of quark flavour mixing shows\cite{data}
that the third family $t$\ and $b$\ quarks are largely decoupled from the
lighter families. At first sight, it looks quite surprising that not only
the 1-3 ``far neighbour'' mixing ($\theta _{13}$) but also the 2-3 ``nearest
neighbour'' mixing ($\theta _{23}$) happen to be small compared with an
``ordinary'' 1-2 Cabibbo mixing ($\theta _{12}$) which is determined,
according to common belief, by the lightest $u$\ and $d$\ quarks. This would
mean that all the other mixings could also be controlled by the lightest
masses $m_{u}$\ and $m_{d}$\ . So, in the chiral symmetry limit $%
m_{u}=m_{d}=0$\ all the flavour mixings disappear.

Recently, in accordance with that view a new mechanism of flavour mixing,
which we called Minimal Flavour Mixing (MFM), was proposed \cite{lfm} (see
the papers in\cite{refs} for further discussion). Remarkably, together with
generically small quark mixing this picture leads to the large neutrino
mixing provided that neutrino masses appear by themselves through a known
''see-saw'' mechanism and the neutrino Dirac and Majorana masses have the
hierarchical structure like as masses of ordinary quarks and leptons.

\section{Where\ does\ flavour\ mixing\ come\ from$?$}

\subsection{Gauge (unrotated) and physical quarks}

Without much loss in generality we could choose the starting mass matrices
of quarks as those being Hermitian and having one texture zero.

\medskip

\noindent{\it 2.1.1~ Upper quarks}

\medskip

\begin{eqnarray}
u^{0},c^{0},t^{0} &\Longleftrightarrow &u,c,t  \nonumber \\
&& \\
\left( 
\begin{array}{lll}
0 & X & Y \\ 
X^{*} & m_{c}^{0} & Z \\ 
Y^{*} & Z^{*} & m_{t}^{0}
\end{array}
\right) &\Leftrightarrow &\left( 
\begin{array}{lll}
m_{u} & 0 & 0 \\ 
0 & m_{c} & 0 \\ 
0 & 0 & m_{t}
\end{array}
\right)  \nonumber
\end{eqnarray}

\ $\ \ \ $

\medskip

\noindent{\it 2.1.2~ Down quarks}

\medskip

\begin{eqnarray}
d^{0},s^{0},b^{0} &\Longleftrightarrow &d,s,b  \nonumber \\
&& \\
\left( 
\begin{array}{lll}
0 & X^{\prime } & Y^{\prime } \\ 
X^{\prime *} & m_{s}^{0} & Z^{\prime } \\ 
Y^{\prime *} & Z^{\prime *} & m_{b}^{0}
\end{array}
\right) &\Longleftrightarrow &\left( 
\begin{array}{lll}
m_{d} & 0 & 0 \\ 
0 & m_{s} & 0 \\ 
0 & 0 & m_{b}
\end{array}
\right)  \nonumber
\end{eqnarray}

\medskip

\noindent{\it 2.1.3~ The measure of Flavour Mixing (in terms of mass
deviations)}

\medskip

\noindent Proposing that the difference between masses of the physical and
gauge quarks of a given generation is much less than the mass of quark of a
previous generation

\vspace{1pt}

\begin{eqnarray}
m_{t}-m_{t}^{0} &\ll &\mbox{ }m_{c},\mbox{ \ \ \ }m_{c}-m_{c}^{0}\ll m_{u}%
\mbox{ \ \ \ }\Longrightarrow \mbox{\ \ \ \ }m_{t}\mbox{\ }-m_{t}^{0}\simeq
m_{u} \\
m_{b}-m_{b}^{0} &\ll &\mbox{ }m_{s},\mbox{ \ \ \ }m_{s}-m_{s}^{0}\ll m_{d}%
\mbox{ \ \ \ }\Longrightarrow \mbox{\ \ \ \ }m_{b}\mbox{\ }-m_{b}^{0}\simeq
m_{d}  \nonumber
\end{eqnarray}
we\vspace{1pt} come to the MFM{\large \ }:

{\it Flavour Mixing is basically determined\ by\ the lightest family mass
generation.} \ \ \ \ {\bf \ \ }

\section{\protect\vspace{1pt} The prototype quark mixing: 2 alternatives}

The proposed MFM driven solely by the generation of the lightest family mass
could actually be realized in two generic ways.

\subsection{Scenario A: ''$u$ and $d$ quark masses running along the
diagonal''}

The first way is when the lightest family mass ($m_{u}$, $m_{d}$ or $m_{e}$)
appears as a result of the complex flavour mixing of all three families. It
``runs along the main diagonal'' of the corresponding $3\times 3$ mass
matrix $M$, from the basic dominant element $M_{33}$ to the element $M_{22}$
(via a rotation in the 2-3 sub-block of $M$) and then to the primordially
texture zero element $M_{11}$ (via a rotation in the 1-2 sub-block). The
direct flavour mixing of the first and third families of quarks and leptons
is supposed to be absent or negligibly small in $M$. Let us note that the
''running along the diagonal'' of the lightest mass means mathematically the
proportionality condition between diagonal and off-diagonal matrix elements
in the $M$ of the type:

\begin{equation}
M_{22}:M_{33}=\left| M_{12}\right| ^{2}:\left| M_{23}\right| ^{2}
\label{pro}
\end{equation}

\medskip

\noindent{\it 3.1.1~ Upper quarks}

\medskip

\noindent Doing so we inescapably come to the mass-matrix for the upper
quarks of the type (using that the trace and determinant of the Hermitian
matrix gives the sum and product of its eigenvalues)

\begin{equation}
M_{13}^{U}=M_{31}^{U}=0\ \ \ \Longrightarrow \left( 
\begin{array}{lll}
0 & \sqrt{m_{u}m_{c}}e^{i\alpha } & 0 \\ 
\sqrt{m_{d}m_{s}}e^{-i\alpha } & m_{c} & \sqrt{m_{u}m_{t}}e^{i\beta } \\ 
0 & \sqrt{m_{u}m_{t}}e^{-i\beta } & m_{t}-m_{u}
\end{array}
\right)
\end{equation}

\vspace{1pt}

\medskip

\noindent{\it 3.1.2~ Down quarks}

\medskip

\noindent Analogously, the mass matrix for the down quarks looks like

\begin{equation}
M_{13}^{D}=M_{31}^{D}=0\Longrightarrow \ \left( 
\begin{array}{lll}
0 & \sqrt{m_{d}m_{s}}e^{i\alpha ^{\prime }} & 0 \\ 
\sqrt{m_{d}m_{s}}e^{-i\alpha ^{\prime }} & m_{s} & \sqrt{m_{d}m_{b}}%
e^{i\beta ^{\prime }} \\ 
0 & \sqrt{m_{d}m_{b}}e^{-i\beta ^{\prime }} & m_{b}-m_{d}
\end{array}
\right)  \label{d0}
\end{equation}

\medskip

\noindent{\it 3.1.3~ Weak mixing \ angles}

\medskip

\noindent So, now all the weak mixing angles can be calculated in terms of
the quark mass ratios and CP-violating phase (which tends to be maximal to
give the right value of the Cabibbo angle\vspace{1pt}) 
\begin{eqnarray}
\left| s_{12}\right| &\simeq &\left| \sqrt{\frac{m_{d}}{m_{s}}}-e^{i(\alpha
-\alpha ^{\prime })}\sqrt{\frac{m_{u}}{m_{c}}}\right| \ \ \ \ \ \ (CP\;phase%
\mbox{ }\delta =\alpha -\alpha ^{\prime }\ tends\ to\mbox{ }\pi /2)\ \ \  
\nonumber \\
\left| s_{23}\right| &=&\left| \sqrt{\frac{m_{d}}{m_{b}}}-e^{i(\beta -\beta
^{\prime })}\sqrt{\frac{m_{u}}{m_{t}}}\right| \ \ \ \ \ \left( 
\begin{array}{c}
0.038(7)\mbox{ }/\mbox{ }0.039(3) \\ 
th\mbox{ }/\exp
\end{array}
\right) \\
\left| \frac{s_{12}}{s_{23}}\right| &=&\sqrt{\frac{m_{u}}{m_{c}}}=0.05(1)\ \
\ \ \ \ \ \ \   \nonumber
\end{eqnarray}

\vspace{1pt}

\subsection{\protect\vspace{1pt}Scenario B: ''$m_{u}$ walking around the
corner, while $m_{d}$ runs along the diagonal''}

The second way, on the contrary, presupposes direct flavour mixing of just
the first and third families. There is no involvement of the second family
in the mixing. In this case, the lightest mass appears in the primordially
texture zero $M_{11}$ element ``walking round the corner'' (via a rotation
in the 1-3 sub-block of the mass matrix $M$). Certainly, this second version
of the MFM mechanism cannot be used for both the up and the down quark
families simultaneously, since mixing with the second family members is a
basic part of the CKM phenomenology (Cabibbo mixing, non-zero $V_{cb}$
element, CP violation). However, this second way could work for the up quark
family provided that the down quarks follow the first way.

\medskip

\noindent{\it 3.2.1~ Upper quarks}

\medskip

\noindent Now we have the new matrix for the upper quarks

\begin{equation}
M_{12}^{U}=M_{21}^{U}=M_{23}^{U}=M_{32}^{U}=0\ \ \ \Longrightarrow \left( 
\begin{array}{lll}
0 & 0 & \sqrt{m_{u}m_{t}}e^{i\alpha } \\ 
0 & m_{c} & 0 \\ 
\sqrt{m_{u}m_{t}}e^{-i\alpha } & 0 & m_{t}-m_{u}
\end{array}
\right)  \label{up}
\end{equation}

\vspace{1pt}

\medskip

\noindent{\it 3.2.2~ Down quarks}

\medskip

\noindent While for the down quarks the above matrix (\ref{d0}) is remained

\begin{equation}
M_{13}^{D}=M_{31}^{D}=0\ \ \ \Longrightarrow \ \left( 
\begin{array}{lll}
0 & \sqrt{m_{d}m_{s}}e^{i\alpha ^{\prime }} & 0 \\ 
\sqrt{m_{d}m_{s}}e^{-i\alpha ^{\prime }} & m_{s} & \sqrt{m_{d}m_{b}}%
e^{i\beta ^{\prime }} \\ 
0 & \sqrt{m_{d}m_{b}}e^{-i\beta ^{\prime }} & m_{b}-m_{d}
\end{array}
\right)  \label{d}
\end{equation}
so that the weak mixing angles and CP-violation phase are given by new
formulas.

\medskip

\noindent{\it 3.2.3~ Weak mixing \ angles}

\medskip

\noindent Actually, the mass matrices (\ref{up}) and (\ref{d}) lead to the
simplest expressions which ever have been derived for \vspace{1pt}\vspace{1pt%
}CKM angles 
\begin{equation}
s_{12}\simeq \sqrt{\frac{m_{d}}{m_{s}}},\ \ \ \ \ s_{23}\simeq \sqrt{\frac{%
m_{d}}{m_{b}}}\ \ \ ,\ \ \ \ s_{13}\simeq \sqrt{\frac{m_{u}}{m_{t}}}\ 
\end{equation}
while the CP violation phase $\delta =\alpha _{U}-\alpha ^{\prime }+\beta
^{\prime }$ is left yet arbitrary.

\subsection{\protect\bigskip \protect\vspace{1pt}CKM matrix}

Our numerical results for both versions of our model, with a maximal CP
violating phase (see discussion in Section 5), are summarized in the
following CKM matrix: 
\begin{equation}
V_{CKM}=\pmatrix{ 0.975(1) & 0.222(4) & 0.0023(5) \ A \cr & & 0.0036(6) \ B
\cr 0.222(4) & 0.975(1) & 0.038(4) \cr 0.009(2) & 0.038(4) & 0.999(1) \cr}
\end{equation}
The uncertainties in brackets are largely given by the uncertainties in the
quark masses (calculated at the electroweak scale\cite{koide}). There is
clearly a real and testable difference between scenarios A and B given by
the value of the $V_{ub}$ element. Everyone can see when looking into
Particle Physics Booklet that the MFM $\ $ansatz \ perfectly \ works{\sl . }
The distinctive predictions for the presently relatively poorly known $V_{ub}
$ and $V_{td}$ elements should be tested in the nearest future. Meanwhile,
the present data from CLEO\cite{cleo} seem to favour scenario B.

\vspace{1pt}

\section{\protect\vspace{1pt}Lepton sector}

The lepton mixing matrix is defined analogously to the CKM matrix: 
\begin{equation}
U=U_{\nu }U{_{E}}^{\dagger }  \label{U}
\end{equation}
where the indices $\nu $ and $E$ stand for $\nu =(\nu _{e},\nu _{\mu },\nu
_{\tau })$ and $E=(e,\mu ,\tau )$. Our model predicts the small charged
lepton mixing angles in the matrix $U_{E}$ . They do not markedly effect
atmospheric neutrino oscillations \cite{superkam}, which appear to require
essentially maximal mixing $\sin ^{2}2\theta _{atm}\simeq 1$. It follows
then that the large neutrino mixing responsible for atmospheric neutrino
oscillations should mainly come from the $U_{\nu }$ matrix associated with
the neutrino mass matrix in (\ref{U}). This requires a different mass matrix
texture for the neutrinos compared to the charged fermions. Remarkably,
there appears to be no need in our case for some different mechanism to
generate the observed mixing pattern of neutrinos: they can get physical
masses and mixings via the usual ``see-saw'' mechanism \cite{GRS}

\begin{equation}
M_{\nu }=-M_{N}^{T}M_{NN}^{-1}M_{N}  \label{nu}
\end{equation}
using the proposed MFM mechanism for their primary Dirac and Majorana
masses, $M_{N}$ and $M_{NN}$, respectively. So, we also have two possible
scenarios in the lepton sector as well.

\vspace{1pt}

\subsection{Scenario A*: ``all the lightest lepton Dirac and Majorana masses
running along the diagonal''}

\vspace{1pt}

\medskip

\noindent{\it 4.1.1~ Charged leptons}

\medskip

\noindent For the charged leptons we have according to the MFM the mixing
angles:

\begin{equation}
\sin \theta _{e\mu }=\sqrt{\frac{m_{e}}{m_{\mu }}}\qquad \sin \theta _{\mu
\tau }=\sqrt{\frac{m_{e}}{m_{\tau }}}\qquad \sin \theta _{e\tau }\simeq 0\ \
\ \ \ \ (in\ U_{CKM}^{lep}=U_{\nu }U{_{E}}^{\dagger })  \label{charged}
\end{equation}
It follows then that the large neutrino mixing responsible for atmospheric
neutrino oscillations should mainly come from the neutrino mass matrix by
itself.

\medskip

\noindent{\it 4.1.2~ Neutrinos}

\medskip

\noindent Let us parametrize the neutrino Dirac mass matrix $M_{N}$ as
follows$\ \ \ $\vspace{1pt} $\ \ \ \ \ \ \ \ \ \ \ \ \ \ \ \ \ $%
\begin{equation}
M_{N}\simeq M_{N3}\left( 
\begin{array}{lll}
0 & y^{3} & 0 \\ 
y^{3} & y^{2} & y^{2} \\ 
0 & y^{2} & 1
\end{array}
\right) \ \ \ \   \label{N}
\end{equation}
proposing the hierarchy like as that for down quarks ($y\approx
(m_{s}/m_{b})^{1/2}\approx 0.15$), while for the Majorana mass matrix $%
M_{NN} $ (for right-handed neutrinos) we take the stronger hierarchy of the
type\cite{fuk}$\ \ \ \ \ \ \ \ \ \ \ \ \ \ \ \ $%
\begin{equation}
M_{NN}\simeq M_{NN3}\left( 
\begin{array}{lll}
0 & y^{5} & 0 \\ 
y^{5} & y^{4} & y^{3} \\ 
0 & y^{3} & 1
\end{array}
\right) \ \ \ 
\end{equation}
which is similar to that for the upper quarks. However, the basic
proportionality condition (\ref{pro}) $\ $underlying the MFM mechanism is
satisfied for both. Now, constructing the mass matrix $M_{\nu }$ (\ref{nu})
for physical neutrinos one immediately comes to the characteristic
predictions for the standard two-flavour atmospheric and solar neutrino
oscillation parameters\cite{fuk} 
\begin{equation}
\sin ^{2}2\theta _{atm}\simeq 1\mbox{ , }\sin ^{2}2\theta _{sun}\simeq \frac{%
2}{3}\mbox{ , }U_{e3}\simeq \frac{1}{2\sqrt{2}}y\mbox{ , }\frac{\Delta
m_{sun}^{2}}{\Delta m_{atm}^{2}}\simeq \frac{1}{4}y^{2}  \label{pred1}
\end{equation}
as compared with the experimentally allowed intervals (for the large-angle
MSW oscillation solution to the solar neutrino problem)

\begin{eqnarray}
0.82 &\leq &\sin ^{2}2\theta _{atm}\leq 1\mbox{ , }0.65\leq \sin ^{2}2\theta
_{sun}\leq 1 \\
\left| U_{e3}\right| ^{2} &\leq &0.05\mbox{ , }5\cdot 10^{-3}\leq \frac{%
\Delta m_{sun}^{2}}{\Delta m_{atm}^{2}}\leq 5\cdot 10^{-2}  \nonumber
\end{eqnarray}

\subsection{\protect\vspace{1pt}\protect\vspace{1pt} Alternative B$^{*}$:
``the lightest Majorana mass walking around the corner,\ while the\ lightest
Dirac masses run along the diagonal''}

$\ $Again, we take for the neutrino Dirac mass matrix\vspace{1pt} $M_{N}$
the form (\ref{N}), while for the Majorana mass matrix $M_{NN}$ we take the
alternative II form

\begin{equation}
M_{NN}\simeq M_{NN3}\left( 
\begin{array}{lll}
0 & 0 & y^{q} \\ 
0 & y^{p} & 0 \\ 
y^{q} & 0 & 1
\end{array}
\right)  \label{m000}
\end{equation}
with an arbitrary eigenvalue hierarchy of the type

\begin{equation}
M_{NN3}:M_{NN2}:M_{NN1}\simeq 1:y^{p}:y^{2q}  \label{h3}
\end{equation}
The seesaw formula (\ref{nu}) then generates an effective physical neutrino
mass matrix $M_{\nu }$ which automatically leads to the large (maximal) $\nu
_{\mu }-\nu _{\tau }$ mixing and small $\nu _{e}-\nu _{\mu }$ mixing for any
hierarchy in the Majorana mass matrix (\ref{m000}) satisfying the condition $%
p\geq 2q-1$ mentioned above. Taking, for an example, $p=5$ and $q=3$ one
naturally comes to the{\sl \ }following predictions for the two flavour
oscillation parameters 
\begin{equation}
\sin ^{2}2\theta _{atm}\simeq 1,\quad \sin ^{2}2\theta _{sun}\simeq \frac{2}{%
9}y^{2},\quad U_{e3}\simeq \frac{1}{\sqrt{2}}y,\quad \frac{\Delta m_{sun}^{2}%
}{\Delta m_{atm}^{2}}\simeq \frac{9}{16}y^{2}  \label{pred2}
\end{equation}
For the known value of the hierarchy parameter $y$ (see above) our
predictions (\ref{pred2}) turn out to be inside of the experimentally
allowed intervals \cite{nu-expSMR} for the small-angle MSW solution for the
solar neutrino oscillation:

\begin{eqnarray}
0.82 &\leq &\sin ^{2}2\theta _{atm}\leq 1,\quad 10^{-3}\leq \sin ^{2}2\theta
_{sun}\leq 10^{-2}  \nonumber \\
\left| U_{e3}\right| ^{2} &\leq &0.05,\quad 5\cdot 10^{-4}\leq \frac{\Delta
m_{sun}^{2}}{\Delta m_{atm}^{2}}\leq 9\cdot 10^{-3}  \label{sma}
\end{eqnarray}
Note that in contrast to the LMA case~(\ref{pred1}), one must include in the
SMA case (\ref{pred2}) even the small contribution stemming from the charged
lepton sector (see Eq.~(\ref{charged}) into the solar neutrino oscillations.

So, one can see that the proposed MFM mechanism works quite successfully in
the lepton sector as well as in the quark sector. Remarkably, the same
mechanism results simultaneously in small quark mixing and large lepton
mixing.

\section{\protect\bigskip Conclusion and outlook}

By its nature, the MFM mechanism is not dependent on the number of
quark-lepton families nor on any ``vertical'' symmetry structure, unifying
quarks and leptons inside a family as in Grand Unified Theories (GUTs). From
the theoretical point of view it is based on the generic proportionality
condition (\ref{pro}) between diagonal and off-diagonal elements of the mass
matrices. For the N family case, this condition could be expressed as: 
\begin{equation}
M_{22}:M_{33}:...:M_{NN}=\left| M_{12}\right| ^{2}:\left| M_{23}\right|
^{2}:...:\left| M_{N-1\ N}\right| ^{2}  \label{prop}
\end{equation}
showing clearly that the heavier families (4th, 5th, ...), had they existed,
would be more and more decoupled from the lighter ones and from each other.
Indeed, this behavior can to some extent actually be seen in the presently
observed CKM matrix elements involving the third family quarks $t$ and $b$.

One might think that the condition ~(\ref{prop}) suggests some underlying
flavour symmetry, probably non-abelian $SU(N)$, treating the N families in a
special way. Indeed, for $N=3$ families, we have found\cite{lfm} that the $%
SU(3)$ chiral family (or horizontal) symmetry \cite{jon}, properly
interpreted in terms of the symmetry breaking vacuum configurations, leads
to the basic condition ~(\ref{pro}) for the mass matrices of the down quarks
and charged leptons leaving the up quark and neutrino Majorana masses to
follow scenario B and B$^{*}$, respectively. One can say that the combined
scenario B+B$^{*}$ for quark and lepton mixing is certainly favored by the $%
SU(3)$ symmetry, since in the $SU(3)$ framework the scenario A wants the up
and down quark mass matrices to be proportional to each other, while
scenario A$^{*}$ requires the exactly lepton mass-like hierarchy for the
Majorana neutrino masses as well, both of which are turned out to be
observationally excluded.

At the same time the symmetry-breaking horizontal scalar fields, triplets
and sextets of $SU(3)$, develop in general complex VEVs and in cases linked
to the MFM mechanism transmit a maximal CP violating phase $\delta =\frac{%
\pi }{2}$ to the effective Yukawa couplings. Apart from the direct
predictability of $\delta $ (which was used in the numerical analysis of the
CKM matrix given in the above), the possibility that CP symmetry is broken
spontaneously like other fundamental symmetries of the Standard Model seems
very attractive---both aesthetically and because it gives some clue to the
flavour part of strong CP violation. On the other hand, spontaneous CP
violation means that the scale of the $SU(3)$ family symmetry must be rather
high (not much less than $M_{GUT}$) in order to avoid the standard domain
wall problem by the well-known inflation mechanism.

So, an $SU(3)$ family symmetry seems to be a good candidate for the basic
theory underlying our proposed MFM mechanism, although we do not exclude the
possibility of other interpretations as well. Certainly, even without a
theoretical derivation of Eq.(\ref{prop}), the MFM mechanism can be
considered as a successful predictive ansatz in its own right. Its further
testing could shed light on the underlying flavour dynamics and the way
towards the final theory of flavour.

We summarize in conclusion the main outputs following from {\it Minimal\
Flavour Mixing}:

\bigskip

$\star ${\large \ }{\it MFM successfully works both in quark and lepton
sector;}

\medskip

$\star \star ${\large \ }{\it MFM can be derived from SU(3) theory of
flavour with extra benefits such as the hierarchy in masses and mixings of
quarks and leptons and the spontaneous (maximal) CP violation;}

\medskip

$\star \star \star ${\large \ }{\it From four yet experimentally allowed
scenarios A+A}$^{*}${\it , A+ B}$^{*}${\it , B+A}$^{*}${\it \ and B+B}$^{*}$%
{\it \ for a combined quark and lepton mixing the scenario B+B}$^{*}${\it \
is most preferable as from the observations, so by an accommodation in SU(3)
theory of flavour;}

\medskip

$\star \star \star \star ${\large \ }{\it While the proposed mechanism
describes well all the presently available data, the most critical
prediction for the \ MFM/B+B}$^{*}${\it \ scenario}$\ ${\it are left to be}$%
\ V_{ub}=\sqrt{\frac{m_{u}}{m_{t}}}${\it \ \ in the quark mixing and SMA
solution to the solar neutrino problem which, if finally confirmed, might
allow to enthusiastically say:}

\bigskip

{\bf ''Minimal Flavour Mixing is in fact Standard Flavour Mixing''.}

\bigskip

\section*{Acknowledgements}

Many thanks to organizers of the SUSY'01 for their impressive work and
warmest hospitality. I also should like to thank B. Arbuzov, H. Fritzsch, D.
Kazakov, A. Kobakhidze, H. Leutwyler, D. Sutherland and Z. -Z. Xing and
especially my collaborators C.Froggatt and H. Nielsen for stimulating
discussions and useful remarks.


\begin{thebibliography}{99}
\bibitem{review}  H. Fritzsch and Z. -Z. Xing, Prog. in Part. and Nucl.
Phys. {\bf 45} (2000) 1.

\bibitem{data}  Particle Data Group, The Europ. Phys. Journ. C {\bf 15}
(2000) 1.

\bibitem{lfm}  J.L. Chkareuli and C.D. Froggatt, Phys.Lett. B {\bf 450}
(1999) 158;

J.L. Chkareuli, C.D. Froggatt and H.B. Nielsen, {\it Minimal Mixing of
Quarks and Leptons in the SU(3) Theory of Flavour, }hep-ph/0109156, Nucl.
Phys. B (to appear).

\bibitem{refs}  K. Matsuda, T. Fukuyama and H. Nishiura, Phys. Rev. D {\bf 60%
} (1999) 013006;

D. Falcone, Mod.Phys.Lett. A {\bf 14} (1999) 1989.

T.K. Kuo, S.W. Mansour and G. -H. Wu, Phys. Rev. D {\bf 60} (1999) 093004;

Shao-Hsuan Chiu, T.K. Kuo and Guo-Hong Wu, Phys. Rev. D {\bf 62} (2000)
053014;

D. Falcone and F. Tramontano, Phys. Rev. D {\bf 63} (2001) 073007;

B.R. Desai and A.R. Vaucher, Phys. Rev. D {\bf 63} (2001) 113001.

\bibitem{koide}  H. Fusaoka and Y. Koide, Phys. Rev. D57 (1998) 3986.

\bibitem{cleo}  CLEO Coll. (B.H. Behrens et al), Phys. Rev. D61 (2000)
052001.

\bibitem{superkam}  Super-Kamiokande Coll. (Y. Fukuda et al), Phys. Rev.
Lett. {\bf 81} (1998) 1562; ibid. {\bf 85 }(2000) 3999.

\bibitem{GRS}  M. Gell--Mann, P. Ramond and R. Slansky, in {\it Supergravity}%
, ed. by P. von Nievenhuizen and D.Z. Friedman, N.Holland, 1979.

\bibitem{fuk}  K. Matsuda, T. Fukuyama and H. Nishiura, see ref.\cite{refs}.

\bibitem{nu-expSMR}  J.N. Bahcall, P.I. Krastev and A.Yu. Smirnov, Phys.
Rev. D {\bf 58} (1998) 096016; J.N. Bahcall, M.C. Gonzales-Garcia and C.
Pena-Garay, hep-ph/0106258.

\bibitem{jon}  J.L. Chkareuli, JETP Lett. {\bf 32} (1980) 671;

Z.G. Berezhiani and J.L. Chkareuli, Yad. Fiz. {\bf 37} (1983) 1043;

F. Wilczek, preprint NSF-ITP-83-08 (1983);

Z.G. Berezhiani, Phys. Lett. B {\bf 129} (1983) 99; ibid B {\bf 150} (1985)
177.

J.C. Wu, Phys. Rev. D {\bf 36} (1987) 1514.

J.L. Chkareuli, Phys. Lett. B {\bf 246} (1990) 498; ibid. B {\bf 300} (1993)
361.
\end{thebibliography}
\end{document}